\documentclass[conference,10pt]{IEEEtran}
\IEEEoverridecommandlockouts
\usepackage{cite}
\usepackage{amsmath,amssymb,amsfonts}
\usepackage{algorithmic}
\usepackage{graphicx}
\usepackage{textcomp}
\usepackage{xcolor}
\usepackage{array}
\usepackage{multirow}
\usepackage{enumitem}
\usepackage{makecell}
\usepackage{float}
\usepackage{subfig}
\usepackage{adjustbox}

\usepackage{caption}
\newcolumntype{B}{>{\raggedright\arraybackslash}m{2 cm}}
\newcolumntype{D}{>{\centering\arraybackslash}m{3 cm}}
\newcolumntype{z}{>{\centering\arraybackslash}m{0.7 cm}}
\newcolumntype{L}{>{\centering\arraybackslash}m{1 cm}}

\def\BibTeX{{\rm B\kern-.05em{\sc i\kern-.025em b}\kern-.08em
		T\kern-.1667em\lower.7ex\hbox{E}\kern-.125emX}}
\usepackage[letterpaper, portrait, margin=0.7in, bmargin=1in]{geometry}

\usepackage{fancyhdr}
\fancypagestyle{firststyle}{
	\fancyhf{}
	\fancyhead[L]{\small Dipankar Shakya, Mingjun Ying, Theodore S. Rappaport, Hitesh Poddar, Peijie Ma, Yanbo Wang, and Idris Al-Wazani, ``Propagation measurements and channel models in Indoor Environment at 6.75 GHz FR1(C) and 16.95 GHz FR3 Upper-mid band Spectrum for 5G and 6G", \textit{(Submitted) 2024 IEEE Global Communications Conference (GLOBECOM), } Cape Town, South Africa, Dec. 2024, pp. 1--6.}
}

\begin{document}
	\bstctlcite{IEEEtran:BSTcontrol}
	\title{Propagation measurements and channel models in Indoor Environment at 6.75 GHz FR1(C) and 16.95 GHz FR3 Upper-mid band Spectrum for 5G and 6G}

	\author{\IEEEauthorblockN{Dipankar Shakya$^{\dagger1}$, Mingjun Ying$^{\dagger2}$, Theodore S. Rappaport$^{\dagger3}$, \\ Hitesh Poddar*, Peijie Ma$^\dagger$, Yanbo Wang$^\dagger$, and Idris Al-Wazani$^\dagger$}
		\IEEEauthorblockA{\textit{$^\dagger$NYU WIRELESS, Tandon School of Engineering, New York University, USA}\\
			\IEEEauthorblockA{\textit{*Sharp Laboratories of America (SLA), Vancouver, Washington, USA}}
			\{$^1$dshakya, $^2$yingmingjun, $^3$tsr\}@nyu.edu}
		\thanks{This research is supported by the New York University (NYU) WIRELESS Industrial Affiliates Program.}
	}
	
	\maketitle
	
	\linespread{1.05}
	
	\thispagestyle{firststyle}
	
	\begin{abstract}
		New spectrum allocations in the 4--8 GHz FR1(C) and 7--24 GHz FR3 mid-band frequency spectrum are being considered for 5G/6G cellular deployments. This paper presents results from the world's first comprehensive indoor hotspot (InH) propagation measurement campaign at 6.75 GHz and 16.95 GHz in the NYU WIRELESS Research Center using a 1 GHz wideband channel sounder system over distances from 11 to 97 m in line-of-sight (LOS) and non-LOS (NLOS). Analysis of directional and omnidirectional path loss (PL) using the close-in free space 1 m reference distance model shows a familiar waveguiding effect in LOS with an omnidirectional path loss exponent (PLE) of 1.40 at 6.75 GHz and 1.32 at 16.95 GHz. Compared to mmWave frequencies, the directional NLOS PLEs are lower at FR3 and FR1(C), while omnidirectional NLOS PLEs are similar, suggesting better propagation distances at lower frequencies for links with omnidirectional antennas at both ends of the links, but also, importantly, showing that higher gain antennas will offer better coverage at higher frequencies when antenna apertures are kept same over all frequencies. Comparison of the omnidirectional and directional RMS delay spread (DS) at FR1(C) and FR3 with mmWave frequencies indicates a clear decrease with increasing frequency. The mean spatial lobe and omnidirectional RMS angular spread (AS) is found to be wider at 6.75 GHz compared to 16.95 GHz indicating more multipath components are found in the azimuthal spatial domain at lower frequencies.
	\end{abstract}

	
	\section{Introduction}
\begin{figure}[htbp]
		\centering
		\includegraphics[width=0.45\textwidth]{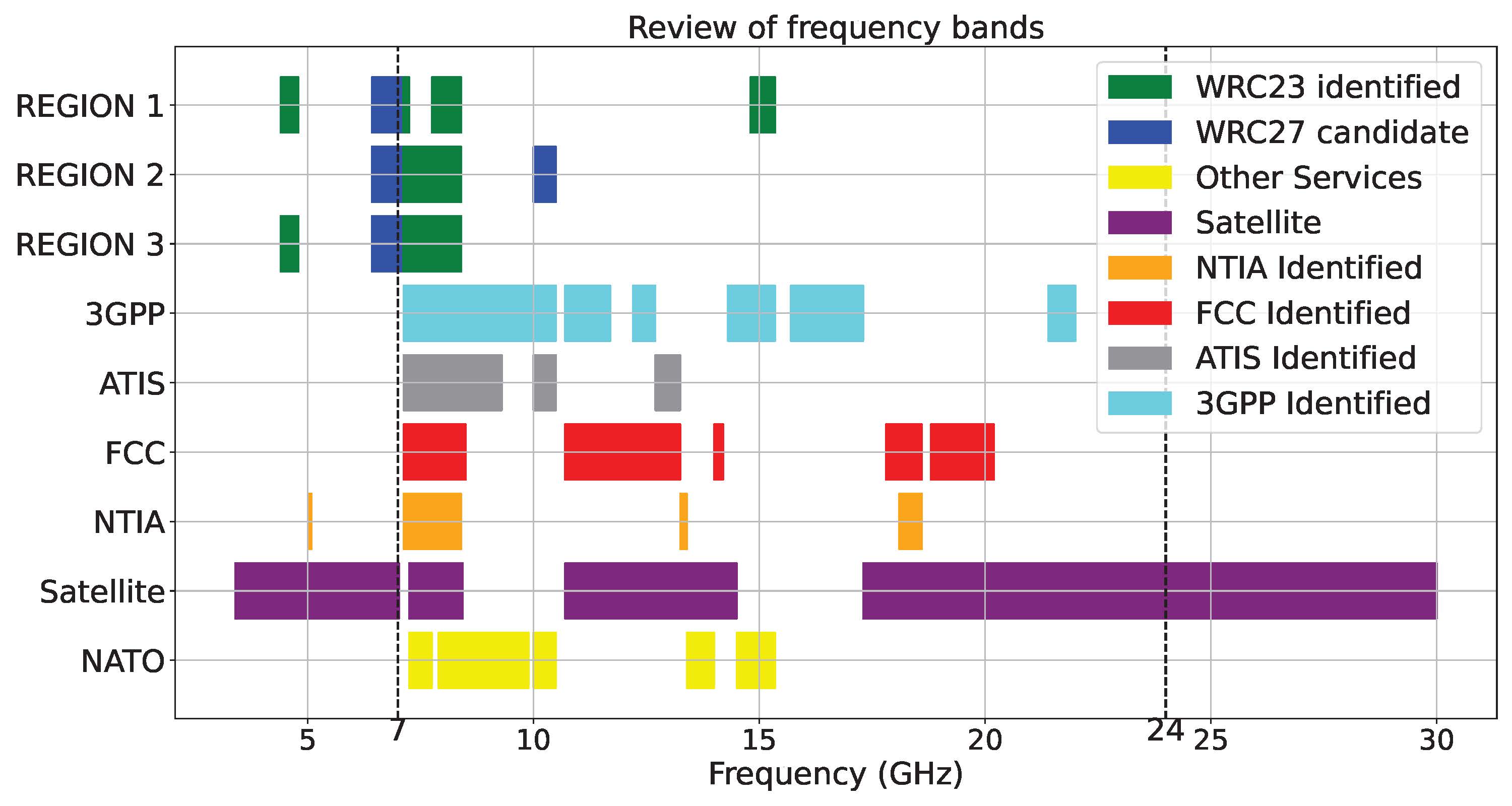}
		\caption{Bands of interest identified in the FR1(C) and FR3 by ITU, NTIA, FCC, and ATIS \cite{NTIA2024,Nokia6GSpectrum, Huawei2021,3GPP3820}.\newline}
		\label{fig:fr3_bands}
		\vspace{-20 pt}
	\end{figure}

	The continued increase in mobile data traffic requires wide bandwidths to deliver gigabits-per-second data throughput for next-generation wireless communication systems. As the spectrum below 6 GHz is congested, the search for higher bandwidth has driven the telecommunications industry towards the large swaths of bandwidth available at higher frequencies for future 5G and 6G deployments. With new spectrum allocations in the 4--8 GHz FR1(C) and 7--24 GHz FR3 frequencies being considered by major government agencies and the International Telecommunication Union (ITU), there is an urgent need for characterizing the radio propagation in the upper mid-band spectrum. The new allocations offer promise of balancing wide area coverage of sub-6 GHz frequencies with higher data rates and capacity of mmWave frequency bands. Telecommunications agencies such as the ITU and the National Telecommunications and Information Administration (NTIA) are exploring effective usage of the FR3 frequencies with other applications such as satellite communications and radio astronomy\cite{NTIA2024,Kang2024OJCOM}. Particularly, ITU and NTIA have focused on frequency bands including 7.125-8.4 GHz, 4.40-4.80 GHz, and 14.8-15.35 GHz, as identified by the ITU World Radio Conference 2023 (WRC-23), Fig. \ref{fig:fr3_bands} \cite{NTIA2024}. There are major efforts from the 3rd Generation Partnership Project (3GPP) and the Federal Communications Commission (FCC) for exploring and standardizing deployments at these frequencies. Industry bodies including the Alliance for Telecommunications Industry Solutions (ATIS) and several mobile operators have identified frequency bands up to 15 GHz for future deployments\cite{Nokia6GSpectrum ,Huawei2021}. There is, therefore, a strong interest and necessity to understand the radio propagation characteristics in the FR3 and FR1(C) frequency bands, including indoor office environments. 
	\vspace{-10 pt} 
	
	Few empirical channel measurements have been reported in the literature in the FR3 and FR1C spectrum. Authors in \cite{zhou2017iet} conducted propagation measurements along an office corridor at 11 and 14 GHz using a vector network analyzer (VNA)-based channel sounder with biconical antennas and observed PLEs of 1.52 and 1.59 in LOS and 3.06 and 2.76 in NLOS. Mean RMS DS were reported as 19.5 ns and 17.9 ns in LOS at 11 GHz and 14 GHz, and 23.43 ns and 22.03 ns at 11 GHz and 14 GHz, respectively. \textit{Wei et al.} conducted propagation measurements in university corridors in \cite{Wei2024vtc} with a MIMO array at 6 GHz with 100 MHz bandwidth using a sliding correlation channel sounder and recorded $\sim$21 ns RMS DS in LOS and $\sim$39 ns in NLOS. Continuous wave measurements conducted along a university corridor in \cite{Oyie2018ia} using a signal generator and analyzer pair at 14 and 22 GHz using horn antennas with 19.5 and 22 dBi gain at the respective frequencies resulted in a directional LOS PLE of 1.6 and 1.7, respectively. Wideband VNA measurements in \cite{janssen1996tc} conducted at 2.4, 4.75, and 11.5 GHz in an indoor office/laboratory environment with biconical antennas yielded LOS PLEs of  1.86, 1.98 and 1.94 and NLOS PLEs of 3.33, 3.75, and 4.46 at the three respective frequencies. Work in \cite{Deng2016gc} measured diffraction loss around drywall, plastic board, and wooden corners, and found good agreement with the Knife Edge Diffraction (KED) model.
	
	The Indoor Hotspot (InH) measurement campaign at the NYU WIRELESS Research Center is the world's first comprehensive radio propagation study at 6.75 GHz in FR1(C) and 16.95 GHz in FR3 bands with a 1 GHz bandwidth channel sounder. The results presented in this paper are analyzed from over 30,000 power delay profiles (PDP) with measurements being conducted at 31 dBm EIRP transmit power employing 15 and 20 dBi gain rotatable horn antennas at 6.75 GHz and 16.95 GHz, respectively, in both co-polarized and cross-polarized antenna configurations. The system and calibration procedures are described in Section \ref{sxn:chSdr} and \ref{sxn:Cal}. The measurement procedure employed is detailed in Section \ref{sxn:EnvProc} with measured path loss models presented in Section \ref{sxn:PL}. Section \ref{sxn:DS} presents spatio-temporal statistics from the measurements before concluding in Section \ref{sxn:Conc}.       
	
	\section{Wideband FR3/FR1C Sliding Correlation Channel Sounder at NYU}
	\label{sxn:chSdr}
	The working principle of the sliding-correlation channel sounder using a pseudo-random noise (PN) sequence is detailed in \cite{Shakya2021tcas,Mac2017jsac}. A 500 Mcps PN sequence, generated at baseband, is upconverted to a 1 GHz wideband signal centered around 6.75 GHz. The FR1(C) channel sounder directly transmits the upconverted signal, whereas the FR3 sounder upconverts the 6.75 GHz IF to a 16.95 GHz RF signal in a heterodyne architecture. At the RX, the received signal is correlated with an identical PN sequence running at a slightly slower rate of 499.9375 Mcps, causing them to slide past one another and resulting in a time-dilated PDP. The work in \cite{Shakya2024ojcom} highlights the FR3/FR1(C) channel sounder with co-located specialized RF frequency converters developed by Mini-Circuits for operation at 6.75 GHz in FR1(C) and 16.95 GHz in FR3. The unique dual-band co-located design allows easy switching of operational frequency bands by changing only the IF signal cable to the suitable RF module input. The detailed specifications of the channel sounder system are provided in Table I in \cite{Shakya2024gc2}.

	\section{Calibration of the Channel Sounder System}
	\label{sxn:Cal}
	The channel sounder calibration for InH measurements ensures accurate capture of the power, delay, and direction of multipath components (MPC). Calibration is conducted at least twice a day--once at the beginning of the measurement day and once at the end--to assure daily and long-term accuracy, and is completed in three distinct phases:
	
	\par\textit{$\bullet$ Linearity and Power Calibration:} The linear range of operation, power limits, and system gain of the RX are defined and confirmed by measurement twice daily, once at the beginning and once at the end of each day. The transmit power (before the TX horn antenna) is measured with a Keysight N1913A power meter via the Keysight N8487A power sensor. The TX attenuator and LO powers are set to keep the TX power within the permissible limit of 35 dBm and ensure linearity. A free space PL measurement at a distance beyond 5$\times$far-field ($D_f$) of the horn antennas helps determine the system gain and the correct attenuator settings. A 4 m separation is selected with a 1.5 m TX and RX height to capture a single LOS path avoiding any reflection paths, adhering to the calibration criteria described in \cite{Xing2018vtc}.   
	
	\par\textit{$\bullet$ Time Calibration:} Accurate measurement of the absolute multipath delay requires correct implementation of time calibration. Before the measurement day, the two Rubidium (Rb) clocks used on the TX and RX are synchronized overnight (12-13 hrs) using a physical cable to ensure accurate time synchronization. During system startup each day, two Raspberry-Pi computers connected to each Rb clock are initiated to run a patent-pending Precision Time Protocol (PTP) synchronization for absolute timing over a Wi-Fi link, as detailed in \cite{Shakya2023gc}. Following the linearity and power calibration at the start of the day, the PDP which captures the single LOS peak is circular shifted to match the free-space propagation delay of 13.3 ns for the four-meter calibration distance. Then, the TX and RX are moved to the desired location for measurements. 
	
	Upon completing measurements at a TX-RX location, the TX and RX are returned to a 4 m separation in the laboratory to remeasure the propagation delay for the LOS multipath and capture any drift in the propagation delay. The drift is caused due to frequency and phase offsets between the untethered Rb clocks at the TX and RX in the duration of the TX and RX measurement, which typically takes several hours due to the intensive sweeping of the channel (described subsequently in Section \ref{sxn:EnvProc}). The drift is found to be unaffected by the movement of TX and RX carts to respective locations\cite{Shakya2023gc}. The observed drift is recorded to correct the measured data to obtain the absolute propagation delay of MPCs. The time calibration is repeated for every TX-RX location measured throughout the day. Unlike past measurement campaigns at NYU WIRELESS, the present campaign uses an active synchronization method that continuously synchronizes the Rb clocks and the PDP propagation delays against a reference MPC during the several hours of measurements at each TX-RX location, using patent-pending methods in \cite{Shakya2023gc}. This new approach removes reliance on ray-tracing or post-processing methods to assure absolute timing of measured PDPs.
	
	
	\par\textit{$\bullet$ Spatial Calibration:} At every TX-RX location, the TX and RX antennas are raised to 2.4 m and 1.5 m above the ground, respectively. The geographical North is used as the spatial 0$^{\circ}$ pointing reference at each TX-RX location pair for consistent AOA/AOD recording across all locations and spatial coordination between TX and RX during measurements.    
	
	Once the calibration (linearity and power, time, and space) process is complete, the channel sounder is ready for multipath propagation measurements at each TX-RX location pair. 
	
	\section{Measurement environment and procedure for InH measurements}
	\label{sxn:EnvProc}
\begin{figure}[htbp]
		\centering
		\includegraphics[width=0.49\textwidth]{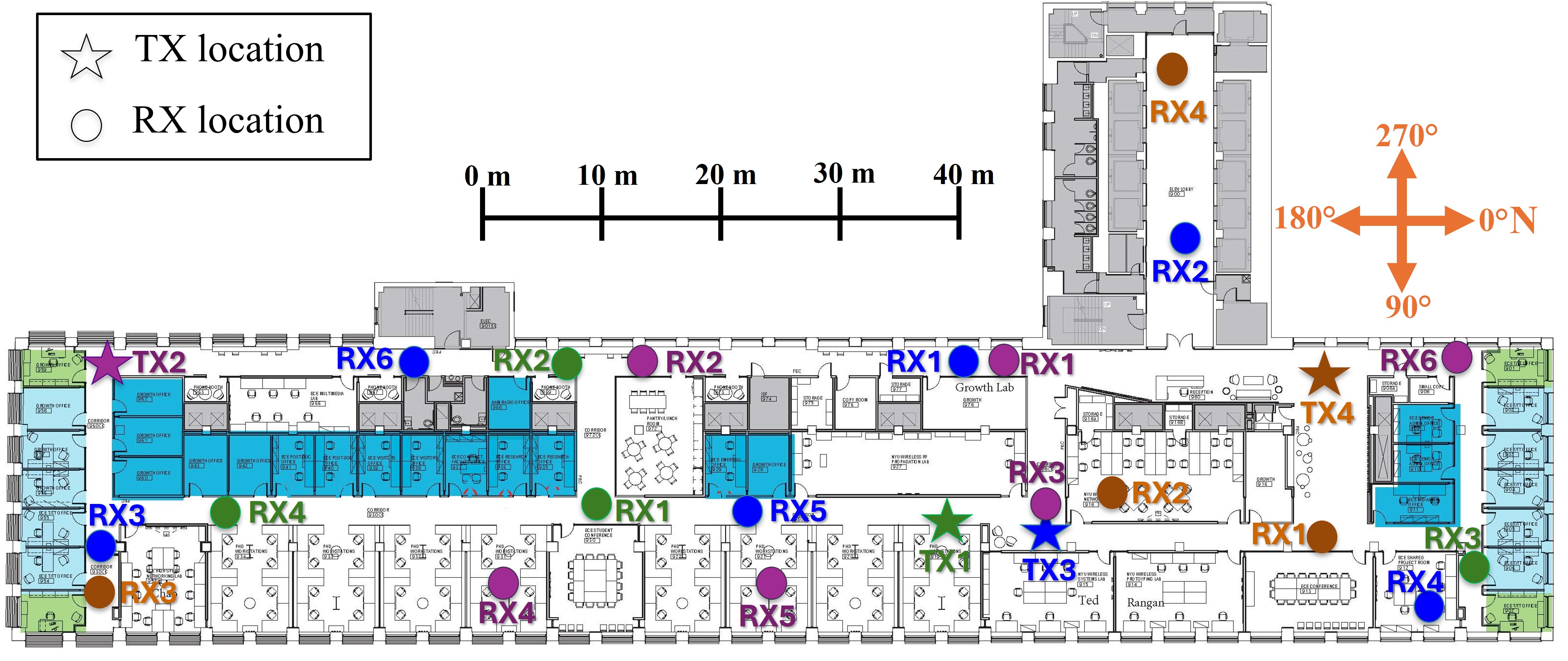}
		\vspace{-10 pt}\caption{The NYU WIRELESS research Center at 370 Jay Street, 9th Floor, Brooklyn, NY is a typical open office space with low height cubicles. The four TX locations are indicated as different color stars with corresponding RX locations as circles of the same color.}
		\label{fig:IndoorMap}
		\vspace{-20 pt}
	\end{figure}

	The InH measurement campaign is conducted in the NYU WIRELESS Research Center at 370 Jay Street, 9th Floor, Brooklyn, NY. The environment is a typical open office space with cubicles, office rooms, labs, and conference rooms. The TX locations are marked as different color stars and RX locations as circles of the corresponding color on the floorplan presented in Fig. \ref{fig:IndoorMap}. Identical T-R locations are measured at both frequency bands. Table \ref{tab:TX-RX} lists the T-R locations in LOS and NLOS for the 6.75 GHz FR1(C) and 16.95 GHz FR3 measurements, encompassing 7 LOS and 13 NLOS T-R location pairs. The T-R separation for the locations range from 11 m to 97 m. No outages are observed at any of the T-R locations in both bands with a transmit EIRP of 31 dBm and link margin (at 5 dB SNR) of 156 dB at 6.75 GHz and 159 dB at 16.95 GHz.
	\vspace{-10 pt}

	\renewcommand{\arraystretch}{1.2}
	\begin{table}[htbp]
		\centering
		\caption{TX-RX location pairs for indoor measurements at 16.95 GHz in FR3 and 6.75 GHz in FR1(C)}
		\begin{tabular}{|r|l|l|}
			\hline
			\multicolumn{3}{|c|}{\textbf{16.95 GHz}} \\
			\hline
			\textbf{TX}    & \textbf{LOS RX} & \textbf{NLOS RX} \\
			\hline
			TX1   & RX1, RX4 & RX2, RX3 \\
			\hline
			TX2   & RX1, RX2, RX6 & RX3, RX4, RX5 \\
			\hline
			TX3   & RX5 & RX1, RX2, RX3, RX4, RX6 \\
			\hline
			TX4   & RX1 & RX2, RX3, RX4 \\
			\hline
			\hline
			\multicolumn{3}{|c|}{\textbf{6.75 GHz}} \\
			\hline
			\textbf{TX}    & \textbf{LOS RX} & \textbf{NLOS RX} \\
			\hline
			TX1   & RX1, RX4 & RX2 \\
			\hline
			TX2   & RX1, RX2 & -- \\
			\hline
			TX3   & RX5 & RX1 RX3, RX4, RX6 \\
			\hline
			TX4   & RX1 & RX2, RX3, RX4 \\
			\hline
		\end{tabular}%
		\label{tab:TX-RX}%
		\vspace*{-1\baselineskip}
	\end{table}%
	\renewcommand{\arraystretch}{1}
	
	\subsection{Measurement Procedure}
	The InH multipath statistics are obtained by capturing the propagation delay, power, AOD, and AOA of all MPCs at each TX-RX location in a two-step process: rapid RX scans, followed by RX azimuthal sweeps in HPBW steps. 
	
	First, the RX completes rapid continuous azimuthal scans (rapid scan) for each TX boresight AOD to rapidly determine the TX AODs with significant energy received at the RX. To initiate the rapid scans, as taught in \cite{Ju2023twc}, the TX and RX are pointed at the direction with strongest received power; boresight in LOS and a strong reflection direction in NLOS. At each TX AOD, the RX performs a continuous 360$^{\circ}$ scan across the azimuth plane. The TX is rotated in increments of HPBW until rapid scans are performed for each AOD. TX AODs are selected using a threshold based on the minimum of two values; either 30 dB below the maximum peak received power among all TX AODs, or 10 dB above the noise floor. All TX AODs with peak power exceeding this threshold are selected for the stepped RX azimuth sweeps.
	
	Once key TX AODs are identified through rapid scan, the TX is pointed to the strongest TX AOD and the RX AOA with the strongest received power is identified. The PDP captured for this AOD-AOA pair is stored and the highest peak in the PDP is marked as the `reference MPC' for implementing the patent-pending PTP synchronization \cite{Shakya2023gc}. After RX completes any azimuthal sweep, the TX and RX return to the AOD-AOA pair of the reference MPC and recapture the PDP to correct any drift in the propagation delay accumulated during an azimuthal sweep to implement successive drift correction (Algorithm I in \cite{Shakya2023gc}). 

	\vspace{-5 pt}
	\renewcommand{\arraystretch}{1.2}
	\begin{table}[htbp]
		\centering
		\caption{TX/RX Elevation angles for different RX azimuthal sweeps for a fixed TX azimuth angle at each TX-RX location}
		\vspace{-5 pt}
		\begin{tabular}{|p{0.75 cm}|p{2.3 cm}|p{4.3 cm}|}
			\hline
			\textbf{Sweep\#} & \textbf{TX elevation} &\textbf{RX elevation} \\
			\hline
			\textbf{1}     & TX is kept at \textit{boresight} elevation& RX is kept at \textit{boresight} elevation. RX is then swept 360$^{\circ}$ in the azimuth plane in HPBW steps. \\
			\hline
			\textbf{2}     & TX is kept at \textit{boresight} elevation& RX is \textit{tilted down by one HPBW} (30$^{\circ}$ at 6.75 GHz/15$^{\circ}$ at 16.95 GHz). RX is then swept 360$^{\circ}$ in the azimuth plane in HPBW steps. \\
			\hline
			\textbf{3}     &TX is kept at \textit{boresight} elevation& RX is \textit{tilted up by one HPBW} (30$^{\circ}$ at 6.75 GHz/15$^{\circ}$ at 16.95 GHz). RX is then swept 360$^{\circ}$ in the azimuth plane in HPBW steps. \\
			\hline
			\textbf{4}     & TX is \textit{tilted down by one HPBW} (30$^{\circ}$ at 6.75 GHz/15$^{\circ}$ at 16.95 GHz).& RX is kept at boresight elevation. RX is then swept 360$^{\circ}$ in the azimuth plane in HPBW steps. \\
			\hline
			\textbf{5}     & TX is \textit{tilted down by one HPBW} (30$^{\circ}$ at 6.75 GHz/15$^{\circ}$ at 16.95 GHz).& RX is \textit{tilted down by HPBW} (30$^{\circ}$ at 6.75 GHz/15$^{\circ}$ at 16.95 GHz). RX is then swept 360$^{\circ}$ in the azimuth plane in HPBW steps. \\
			\hline
		\end{tabular}%
		\label{tab:sweeps}%
		\vspace*{-1\baselineskip}
	\end{table}%
	\renewcommand{\arraystretch}{1.0}
	 
	With key TX AODs and the reference MPC identified, the RX azimuthal sweeps, where the RX is rotated 360 $^{\circ}$ in the azimuth in HPBW increments, are initiated to capture multipath in the channel. Five different RX azimuthal sweeps, as shown in Table \ref{tab:sweeps} are completed for each TX AOD. With the TX and RX at boresight elevation, the RX steps through the azimuth in HPBW steps, referred to as Sweep\#1 in Table \ref{tab:sweeps}. At each AOA, during the azimuthal scans, the RX averages 20 instantaneous PDPs and logs the system and spatial information corresponding to the captured directional PDP. In the 16.95 GHz FR3 campaign, an RX azimuthal sweep includes 24 AOAs corresponding to the 15$^{\circ}$ HPBW (15$^{\circ}$$\times$24 = 360$^{\circ}$), while the 6.75 GHz FR1(C) campaign includes 12 AOAs using a 30$^{\circ}$ HPBW (30$^{\circ}$$\times$12 = 360$^{\circ}$). Omnidirectional PDPs are synthesized from directional PDPs using the method described in \cite{Sun2015gc} despite different antenna beamwidths to obtain omnidirectional channel statistics.
	
	Following Sweep\#1, for the same particular TX AOD, the RX elevation is down- and up- tilted by HPBW for sequential azimuthal sweeps (Sweep\#2 and Sweep\#3 in Table \ref{tab:sweeps}). For Sweep\#4, the TX elevation is downtilted by HPBW and the stepped sweep is performed with RX at boresight elevation. Then, Sweep\#5 is conducted with both TX and RX downtilted by HPBW. The TX then moves to the next identified TX AOD and the five RX azimuthal sweeps are repeated until all identified TX AODs are measured. After the TX steps through all identified AODs, the RX azimuthal sweeps are repeated for a cross-polarized (V-H) antenna configuration, whereby the TX horn antenna is oriented in V-polarization and the RX horn antenna in H-polarization. Using the measured cross-polarization discrimination of 35.7 and 38.4 dB at 6.75 GHz and 16.95 GHz respectively \cite{Shakya2024gc2}, only the TX AODs that exhibited a peak received power in co-polarized (V-V) configuration of 30 dB above the noise floor during rapid scans are measured. The attenuation at RX must be lowered to capture the much weaker multipath in the V-H configuration due to increased cross-polarization propagation loss\cite{Xing2019gc}.  
	
\begin{figure}
		\centering%
		\subfloat[]{%
			\centering
			\includegraphics[width=40mm]{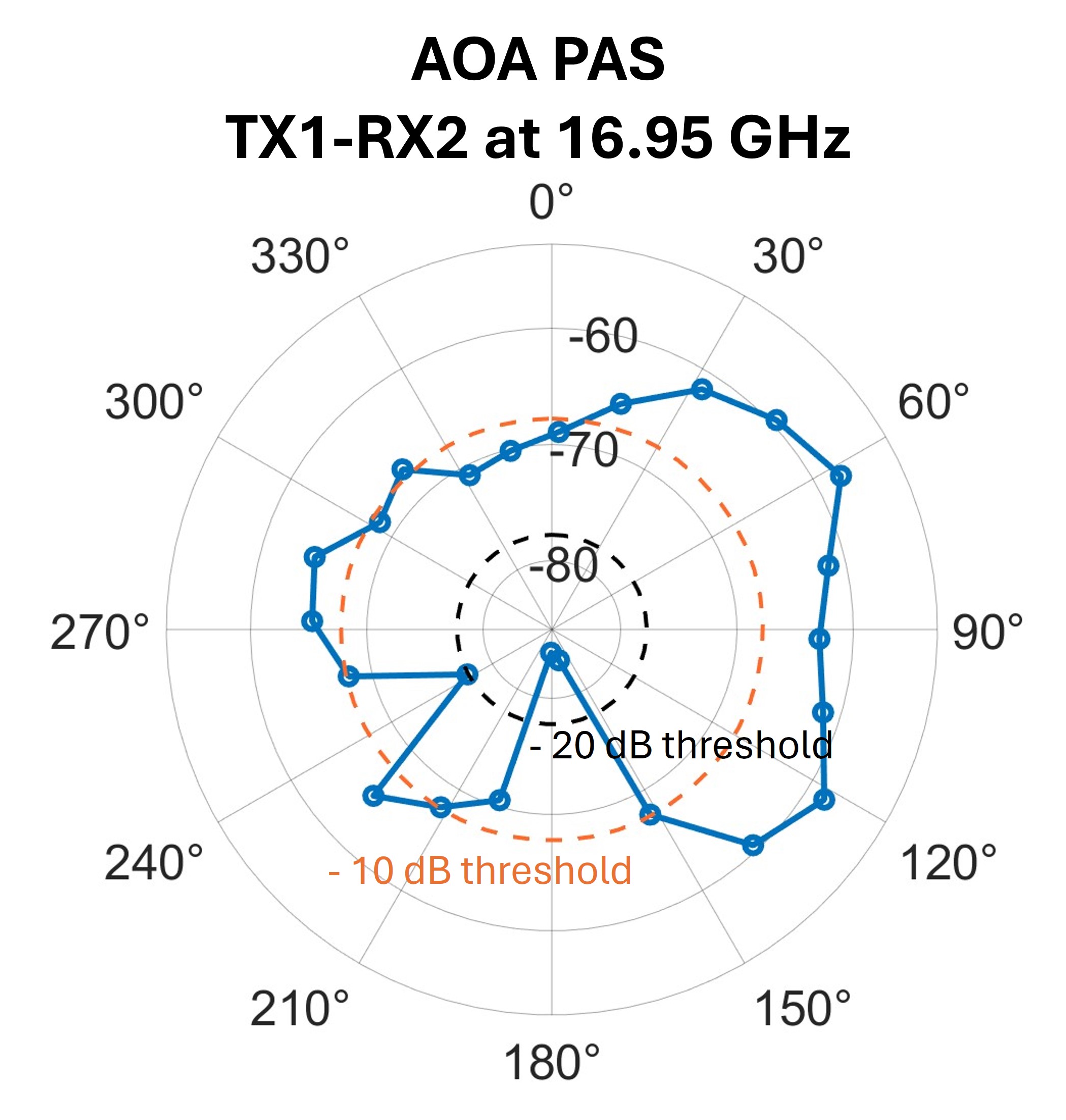}
		}%
		\subfloat[]{%
			\centering
			\includegraphics[width=37mm]{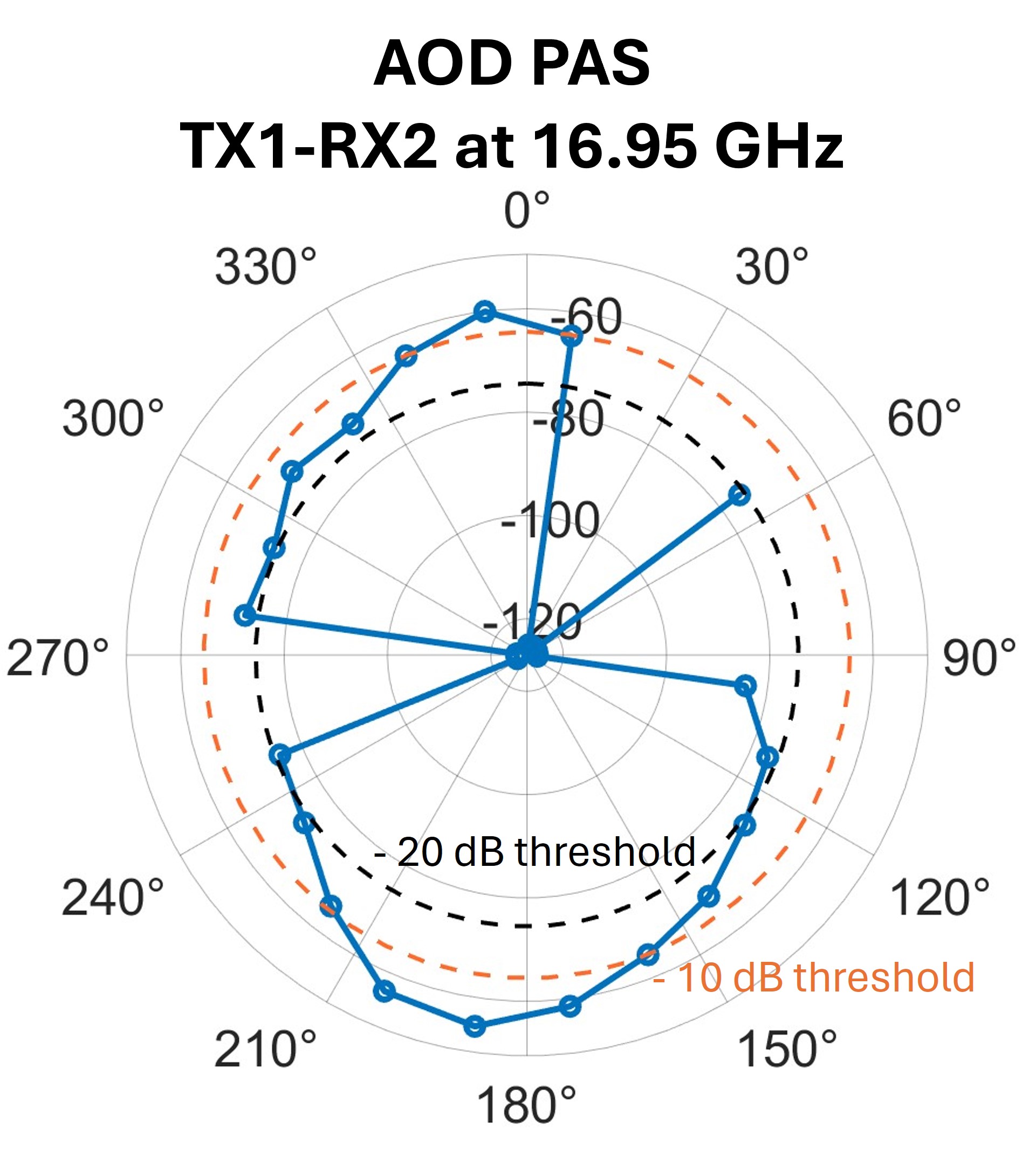}
		}%
		\\
		\caption{Power Azimuth Spectrum for TX1-RX2 in the 16.95 GHz campaign at: (a) RX AOAs (b) TX AODs. [T-R separation: 27 m]}
		\label{fig:PAS}
		\vspace{-20 pt}
	\end{figure}

	After collecting directional PDPs from RX azimuthal sweeps over all key TX AODs at a T-R location pair, post processing on measured PDPs is conducted by applying a power threshold to each directional PDP. The higher received power of either 5 dB above the noise floor or 25 dB below the PDP peak is selected as the threshold and powers below it are ignored in the processing of channel statistics (Section IV B in \cite{Shakya2024TAP}). The power azimuth spectrum (PAS) generated from the directional PDPs at each T-R location produced wide spatial lobes with power received at several contiguous HPBW pointing directions. A spatial lobe is defined as contiguous pointing directions with received power exceeding a spatial lobe threshold (SLT) in the PAS, below which powers in the PAS are ignored for calculating angular statistics such as RMS AS\cite{Shakya2024TAP}. An SLT of 10 dB below the peak received power in the PAS is used for defining the spatial lobes in the InH campaigns, shown in Fig. \ref{fig:PAS} as the orange dotted circle. An example of a PAS recorded for different TX AOD and RX AOA in the 16.95 GHz campaign for the TX1-RX2 pair (Fig. \ref{fig:IndoorMap}) is depicted in Fig. \ref{fig:PAS}.  
	
	\section{Large Scale Path Loss Modeling}
	\label{sxn:PL}    
	The close-in 1 m free space reference distance (CI) model, \eqref{eq:CI}, for describing the path loss is employed on the directional and omnidirectional path loss measurements. The CI model efficiently describes PL with a single PLE (n), as demonstrated in Appendix A \cite{maccartney2015ia}.
	\begin{equation}
		\label{eq:CI}
		\begin{split}
			PL^{CI}(f_c,d_{\text{3D}})\;\text{[dB]} &= \text{FSPL}(f_c, 1 m)+\\&10n\log_{10}\left( \dfrac{d_{3D}}{d_{0}} \right)+\chi_{\sigma},\\
			\text{FSPL}(f_c,1 m) &= 32.4 + 	20\log_{10}\left(\dfrac{f_c}{1\;\text{GHz}}\right),
		\end{split}
	\end{equation}
	where FSPL$(f_c, 1 \;\text{m})$ is obtained for carrier frequency $f_c$ GHz at 1 m, $n$ is the PLE, and $\chi_{\sigma}$ is large-scale shadow fading (zero-mean Gaussian r.v. with s.d. $\sigma^{CI}$ in dB) \cite{Rappaport2015tc}.
	
	\renewcommand{\arraystretch}{1.2}
	\begin{table*}[t]
		\centering
		\caption{Directional and Omnidirectional CI PL model parameters for 6.75 and 16.95 GHz FR1(C) and FR3 InH campaigns, with comparison at 28, 73 and 142 GHz FR2 campaigns}\vspace{-1 em}
		\begin{tabular}{|@{}D@{}|L|L|z|z|z|z|z|z|z|z|z|z|}
			\cline{1-13}
			\multirow{3}{2.1 cm}{\vfil \centering \textbf{Campaign}} & \multirow{3}{1 cm}{\vfil \centering \textbf{Distance (m)}} & \multirow{3}{1 cm}{\centering \textbf{Antenna HPBW (TX/RX)}} & \multicolumn{6}{c|}{\textbf{Directional path loss}} & \multicolumn{4}{c|}{\textbf{Omni path loss}} \\
			\cline{4-13}    \multicolumn{1}{|c|}{} & \multicolumn{1}{c|}{ } & \multicolumn{1}{c|}{ } & \multicolumn{2}{c|}{LOS} & \multicolumn{2}{c|}{NLOS Best} & \multicolumn{2}{c|}{NLOS} & \multicolumn{2}{c|}{LOS} & \multicolumn{2}{c|}{NLOS}  \\
			\cline{4-13}    \multicolumn{1}{|c|}{} & \multicolumn{1}{c|}{} & \multicolumn{1}{c|}{} & \multicolumn{1}{z|}{n} & \multicolumn{1}{z|}{$\sigma$ (dB)} & \multicolumn{1}{z|}{n} & \multicolumn{1}{z|}{$\sigma$ (dB)} & \multicolumn{1}{z|}{n} & \multicolumn{1}{z|}{$\sigma$ (dB)} & \multicolumn{1}{z|}{n} & \multicolumn{1}{z|}{$\sigma$ (dB)} & \multicolumn{1}{z|}{n} & \multicolumn{1}{z|}{$\sigma$ (dB)}\\
			\cline{1-13}
			\centering \textbf{6.75 GHz (This work)} & 11-97 & (30$^{\circ}$/30$^{\circ}$) & 1.55  & 2.52  & 2.74  & 7.14  & 3.05  & 9.71  & 1.40  & 3.41  & 2.42  & 7.87 \\
			\cline{1-13}
			\centering \textbf{16.95 GHz (This work)} & 11-97  & (15$^{\circ}$/15$^{\circ}$) & 1.45  & 1.87  & 3.52  & 9.28  & 3.93  & 14.90 & 1.32  & 2.66  & 3.07  & 9.03 \\
			\cline{1-13}
			\centering \textbf{28 GHz \cite{Ju2021jsac}} & 4-46  & (30$^{\circ}$/30$^{\circ}$) & 1.70  & 2.90  & 3.30  & 10.80  & 4.4  & 12.10 & 1.2  & 1.80  & 2.7  & 9.70 \\
			\cline{1-13}
			\centering \textbf{73 GHz \cite{Xing2021_Inicl}} & 4-46  & (15$^{\circ}$/15$^{\circ}$) & 1.63  & 3.06  & 3.30  & 8.76  & 5.51  & 8.94 & 1.36  & 2.30  & 2.81  & 8.71 \\
			\cline{1-13}
			\centering \textbf{142 GHz \cite{Xing2021_Inicl}} & 4-39  & (8$^{\circ}$/8$^{\circ}$) & 2.05  & 2.89  & 3.21  & 6.03  & 4.60  & 13.80 & 1.74  & 3.62  & 2.83  & 6.07 \\
			\cline{1-13}
		\end{tabular}%
		\label{tab:PLEs}%
		\vspace*{-1\baselineskip}
		\vspace{-10 pt}
	\end{table*}%
	\renewcommand{\arraystretch}{1} 
	
\begin{figure}
		\centering%
		\subfloat[]{%
			\centering
			\includegraphics[width=45mm]{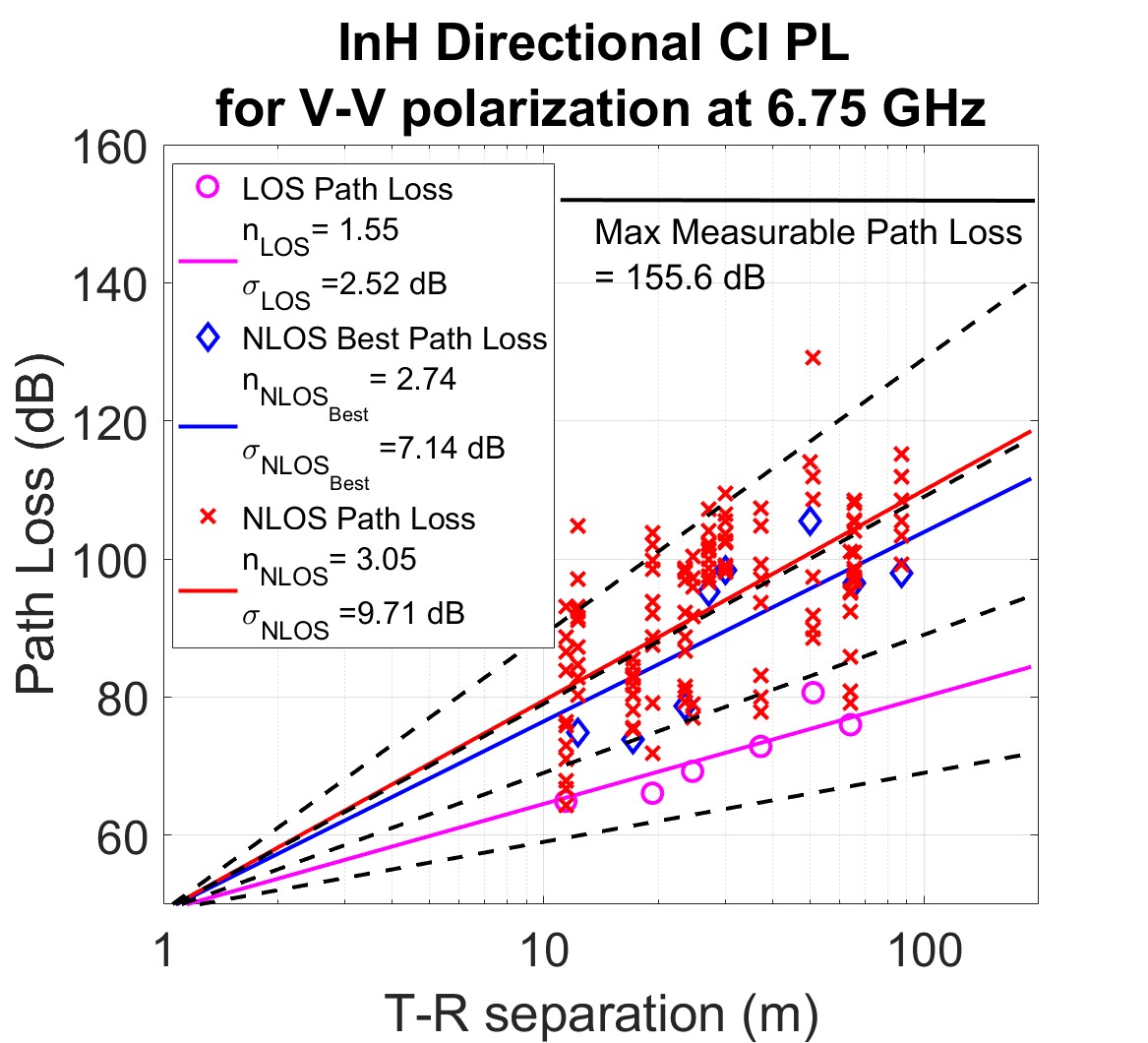}
			\label{fig:dirpl7}
		}%
		\subfloat[]{%
			\centering
			\includegraphics[width=45mm]{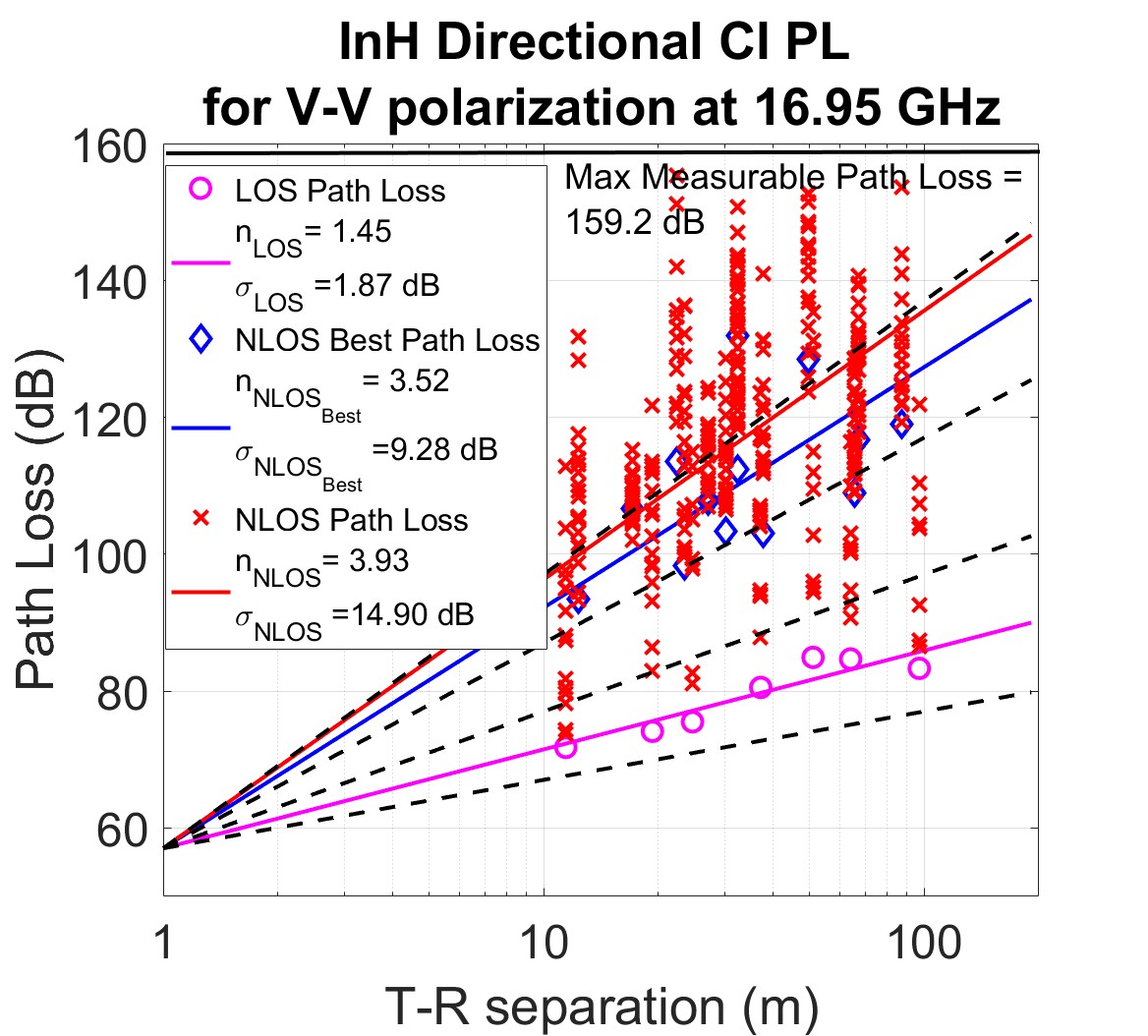}
			\label{fig:dirpl17}
		}%
		\\[2.6mm]
		\vspace{-10 pt}
		\caption{InH directional CI PL models and scatter plot for V-V polarization at: (a) 6.75 GHz FR1(C); (b) 16.95 GHz FR3. [T-R separation: 11--97 m]}
		\vspace{-20 pt}
		\label{fig:dirpl}
	\end{figure}

	\textit{Directional PL modeling} results are based on the definitions for LOS, NLOS$_{\text{Best}}$, NLOS presented in Table IV in \cite{Rappaport2015tc}. 
	The scatter plot of recorded PL in V-V configuration and the CI PL models are presented in Fig. \ref{fig:dirpl} 
	
	\textit{Omnidirectional PL modeling} is carried out by superposition of the directional received powers at each unique non-overlapping azimuth and elevation pointing angles removing the antenna gain, following the method given in \cite{Sun2015gc}.
	
	The CI PL model results for directional and omnidirectional PL, removing the antenna gains, are tabulated in Table \ref{tab:PLEs} along with results from past measurement campaigns in indoor offices at mmWave (FR2) frequencies.
	
	\begin{figure}
		\centering%
		\subfloat[]{%
			\centering
			\includegraphics[width=46mm]{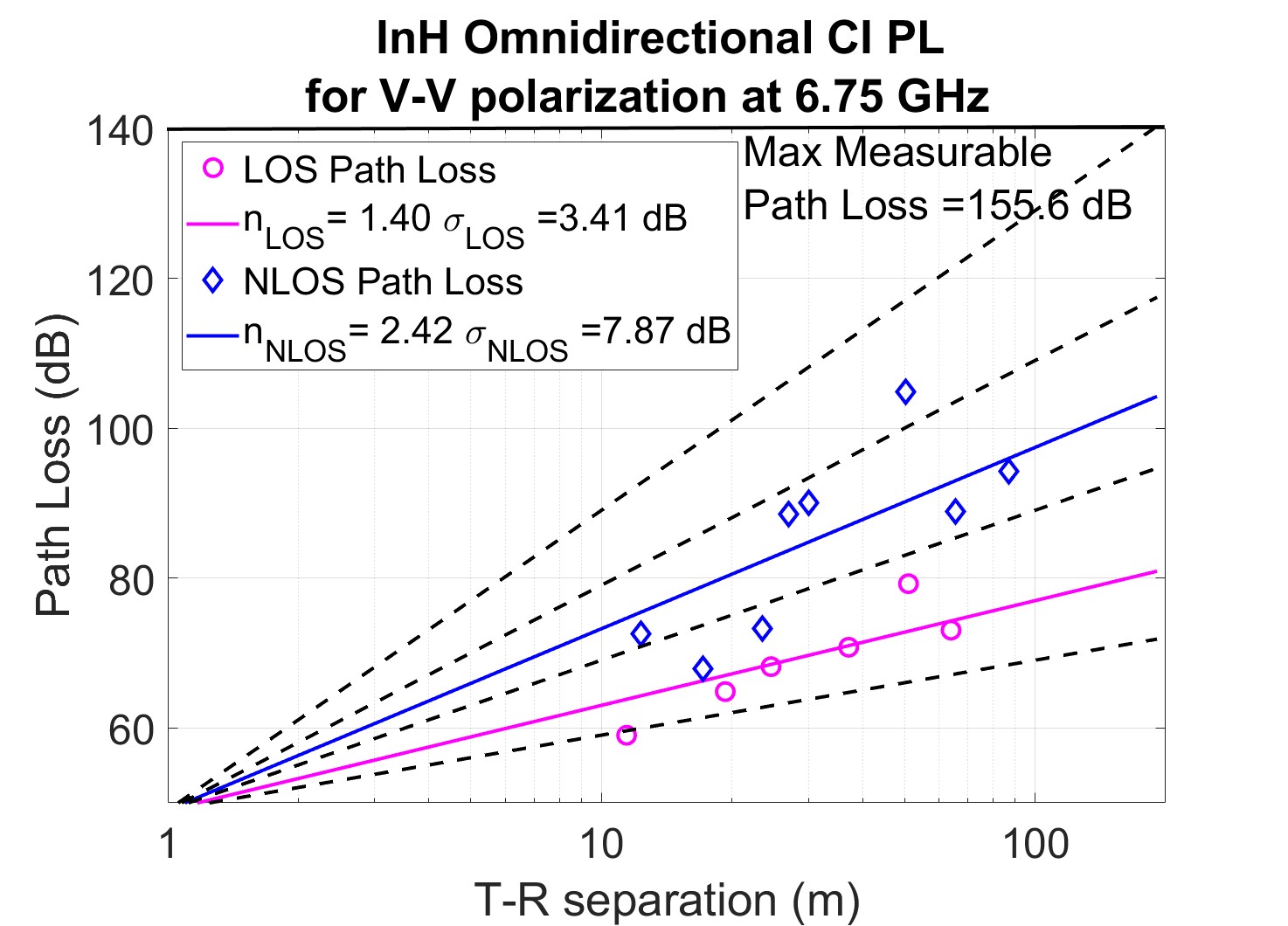}
			\label{fig:omnipl7}
		}%
		\subfloat[]{%
			\centering
			\includegraphics[width=46mm]{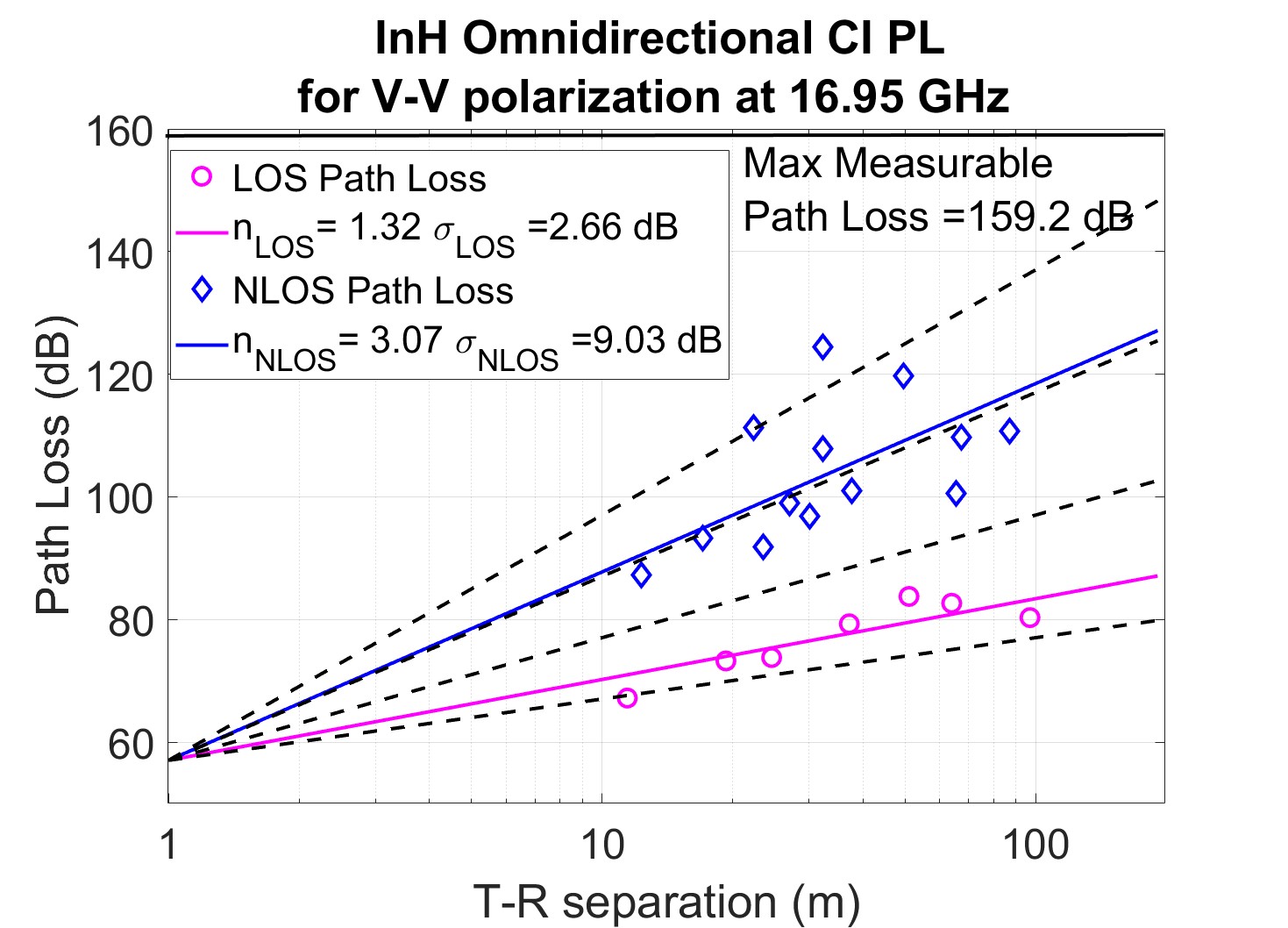}
			\label{fig:omnipl17}
		}%
		\\[2.6mm]
		\vspace{-10 pt}
		\caption{InH omnidirectional CI PL models and scatter plot for V-V polarization at: (a) 6.75 GHz FR1(C); (b) 16.95 GHz FR3. [T-R separation: 11--97 m]}
		\vspace{-20 pt}
	\end{figure}
	
	The LOS directional PL exponent (PLE) is obtained as 1.55 at 6.75 GHz and 1.45 at 16.95 GHz, which are slightly lower (i.e. the channel is less lossy) than the directional LOS PLEs at 28, 73, and 142 GHz (1.7, 1.63, and 2.05, respectively). Further, the NLOS directional PLE at 6.75 and 16.95 GHz are 3.05 and 3.93, which are significantly lower than the PLEs at 28, 73, and 142 GHz (4.4, 5.51, and 4.6, respectively). The NLOS$_{\text{Best}}$ PLEs are found to be 2.74 and 3.52 at 6.75 and 16.95 GHz, respectively, which are comparable to higher mmW channels. The omnidirectional PL modeling resulted in LOS PLEs of 1.4 at 6.75 GHz and 1.32 at 16.95 GHz. Similarly, NLOS PLEs of 2.42 and 3.07 are obtained at 6.75 GHz and 16.95 GHz, respectively, which are similar to the PLE values at the mmwave frequencies. 
	
	\section{Spatio-Temporal InH Statistics}
	\label{sxn:DS}
	The RMS DS is a measure of the temporal dispersion of multipath in a wideband channel. The power threshold of 25 dB below PDP maxima or 5 dB above noise floor, discussed in Section \ref{sxn:EnvProc}, is used to calculate RMS DS. The mean of the RMS DS observed in the directional PDPs captured in the 6.75 GHz campaign is obtained as 19.3 ns and 21.7 ns in LOS and NLOS, whereas, the directional RMS DS from the 16.95 GHz measurements are obtained as 19.5 ns and 17.01 ns.  After synthesizing the omnidirectional PDP for each TX-RX location \cite{Sun2015gc}, the RMS DS in the omnidirectional PDP is observed as 33.7 ns and 43.5 ns in LOS and NLOS at 6.75 GHz, and 22.1 ns and 40.7 ns in LOS and NLOS at 16.95 GHz. It is clear that the 6.75 GHz band has slightly greater RMS DS compared to the 16.95 GHz band.
	
	Further, Table \ref{tab:RMS_DS} compares the RMS DS with results at higher frequencies. The RMS DS experiences a clear decreasing trend with increasing frequency, which is evident across both directional and omnidirectional RMS DS values in LOS and NLOS conditions. 
	
\begin{figure}[!t]
		\centering
		\subfloat[]{%
			\includegraphics[width=0.48\columnwidth]{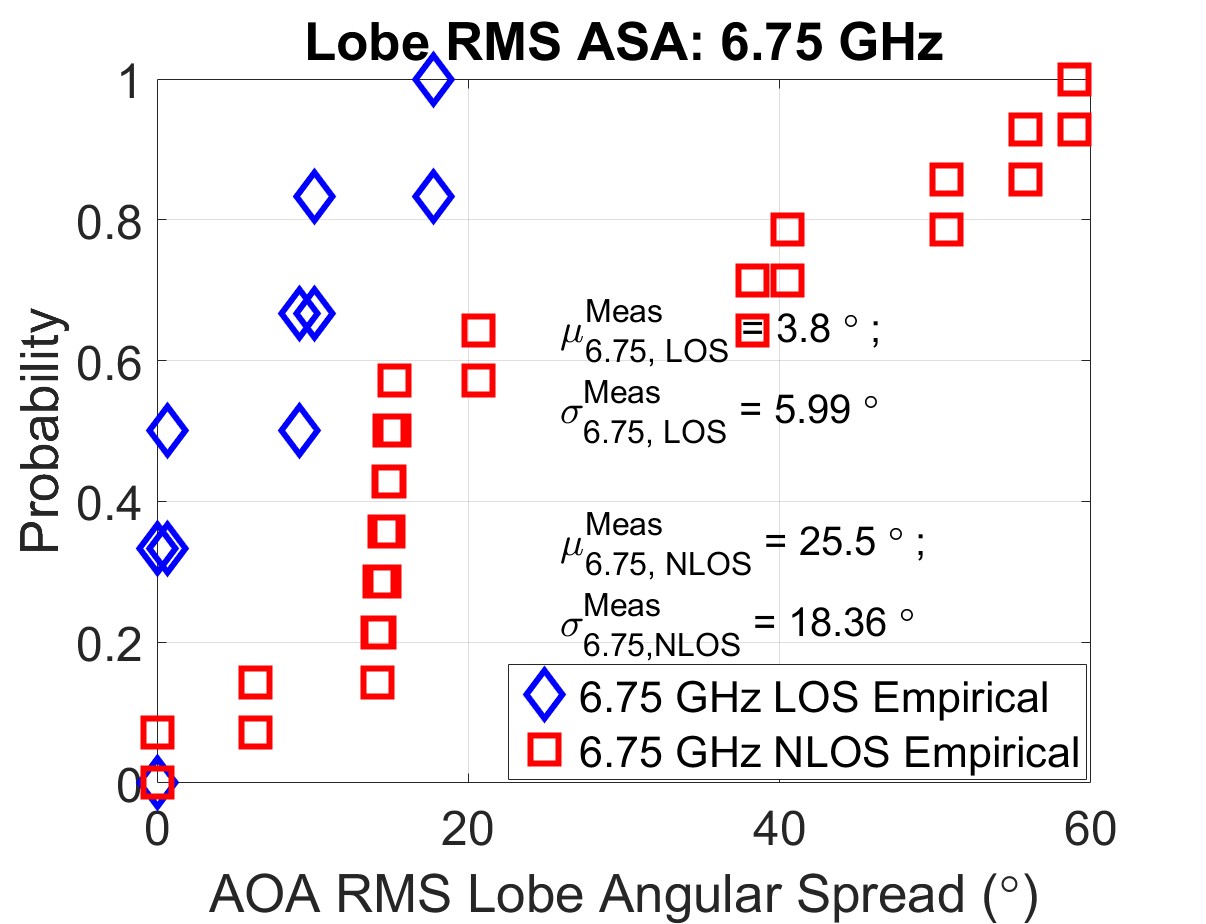}
			\label{fig:diras7}
		}%
		\hfil 
		\subfloat[]{%
			\includegraphics[width=0.48\columnwidth]{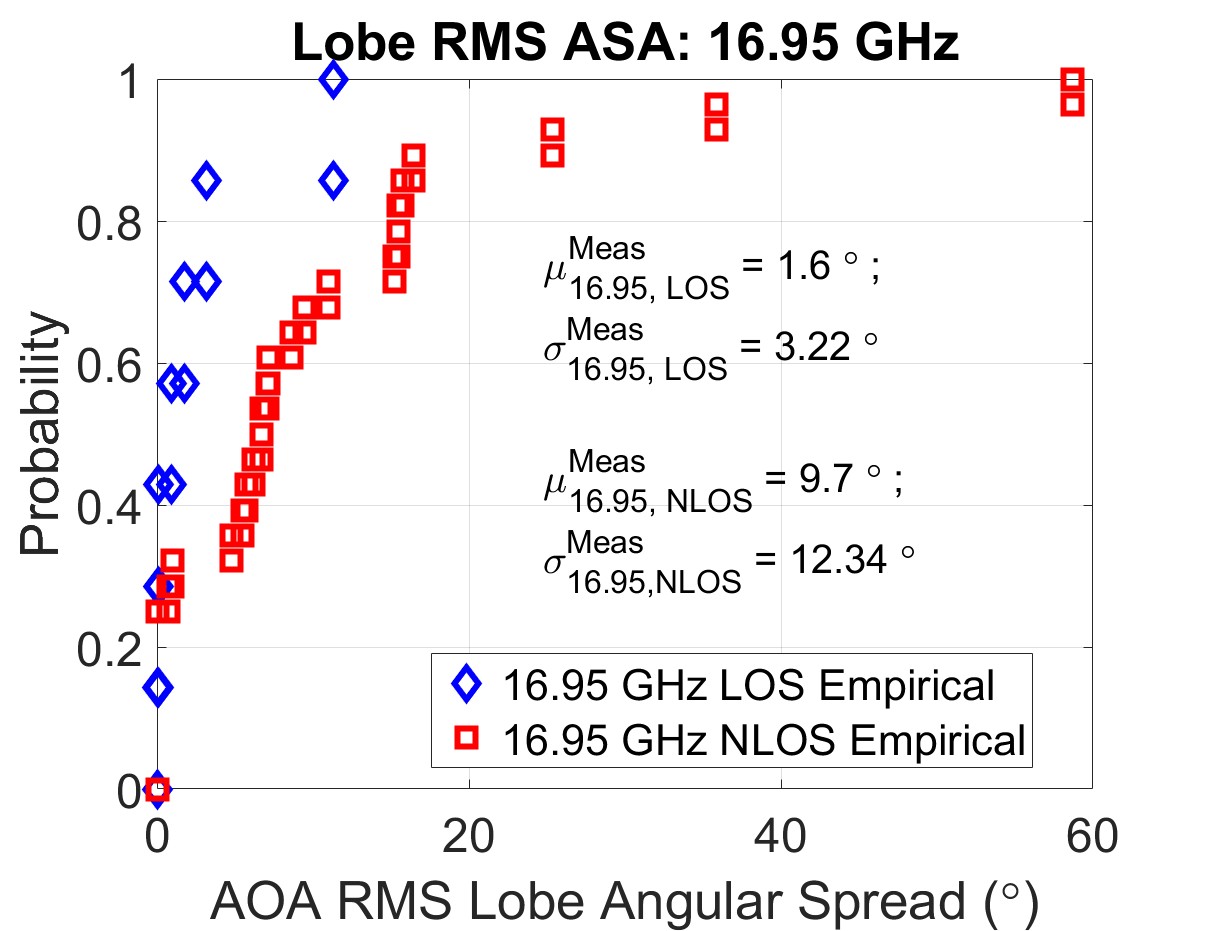}
			\label{fig:diras17}
		}%
		\\ 
		\vspace{-10 pt}
		\subfloat[]{%
			\includegraphics[width=0.48\columnwidth]{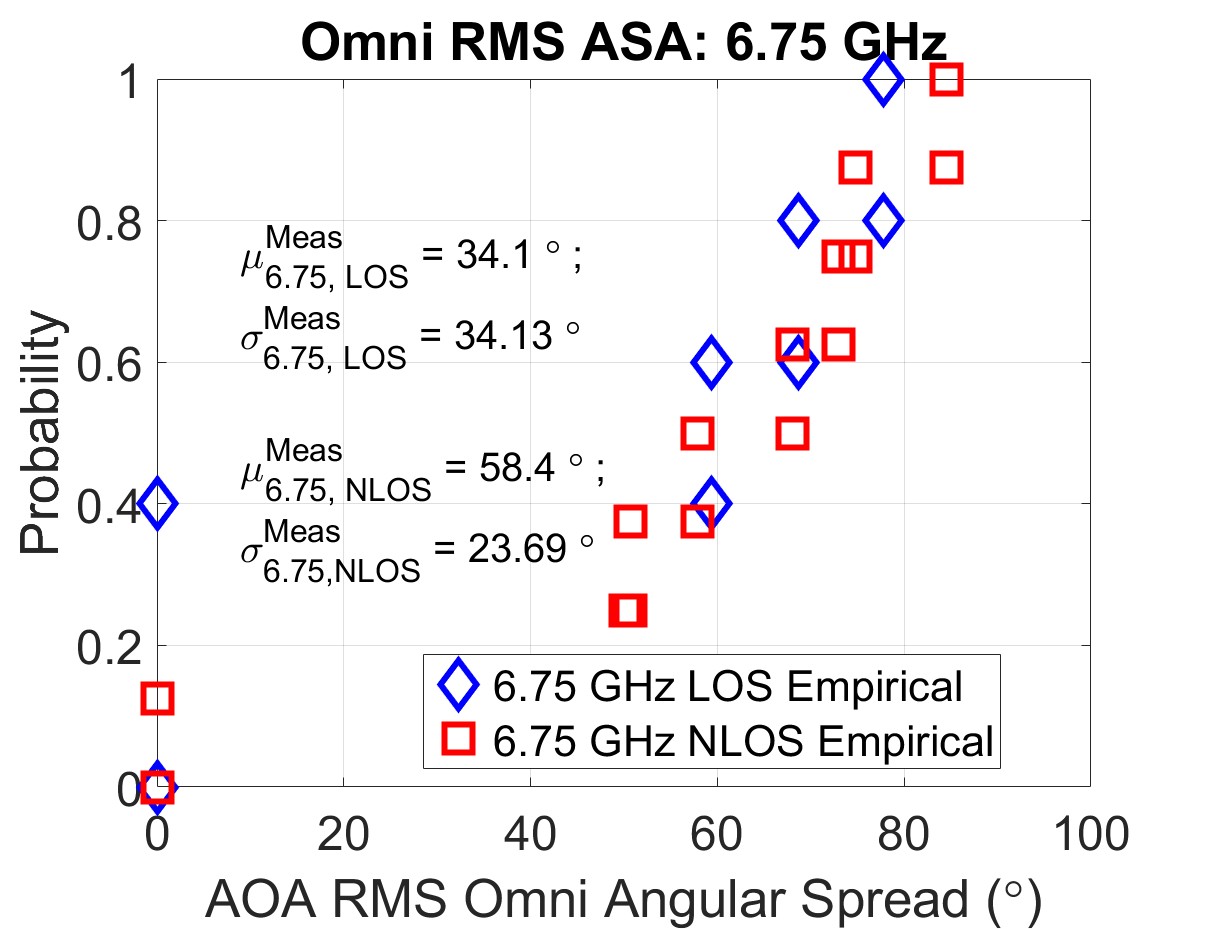}
			\label{fig:omnias7}
		}%
		\hfil 
		\subfloat[]{%
			\includegraphics[width=0.48\columnwidth]{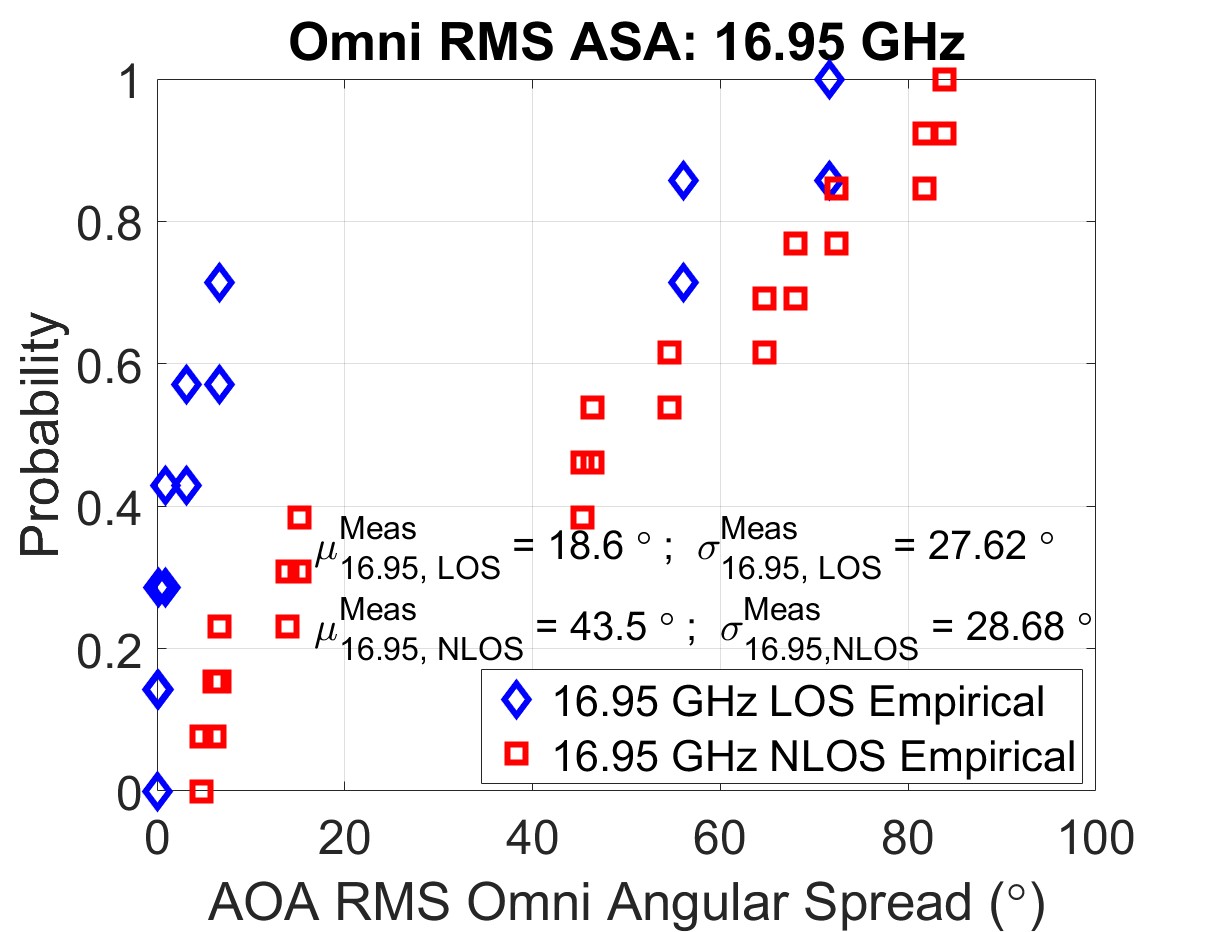}
			\label{fig:omnias17}
		}%
		\caption{InH LOS and NLOS RMS AS for AOA PASs at different frequencies: (a) Lobe ASA at 6.75 GHz; (b) Lobe ASA at 16.95 GHz; (c) Omni ASA at 6.75 GHz; (d) Omni ASA at 16.95 GHz. }
		\label{fig:2x2ASfig}
		\vspace{-15 pt}
	\end{figure}

	The RMS AS captures the azimuthal angular dispersion in the PAS, as shown in Fig. \ref{fig:PAS}. Fig. \ref{fig:2x2ASfig} shows the CDF of the lobe and omni AS for the AOA PAS at 6.75 and 16.95 GHz for both LOS and NLOS scenarios. The mean AS is found to be higher in NLOS than LOS for both lobe and omni AS, which has also been reported at higher frequencies\cite{Ju2021jsac,Xing2021_Inicl}. Omni AOA AS (ASA) are observed at 34$^{\circ}$ in LOS and 58$^{\circ}$ in NLOS at 6.75 GHz, whereas they are found lower at 16.95 GHz at 18$^{\circ}$ in LOS and 43$^{\circ}$ in NLOS. This further confirms less multipath dispersion in both time and space at higher mmWave frequencies, as seen in \cite{Xing2021_Inicl,Ju2021jsac}.

	\begin{figure}[!t]
		\centering
		\subfloat[]{%
			\includegraphics[width=0.48\columnwidth]{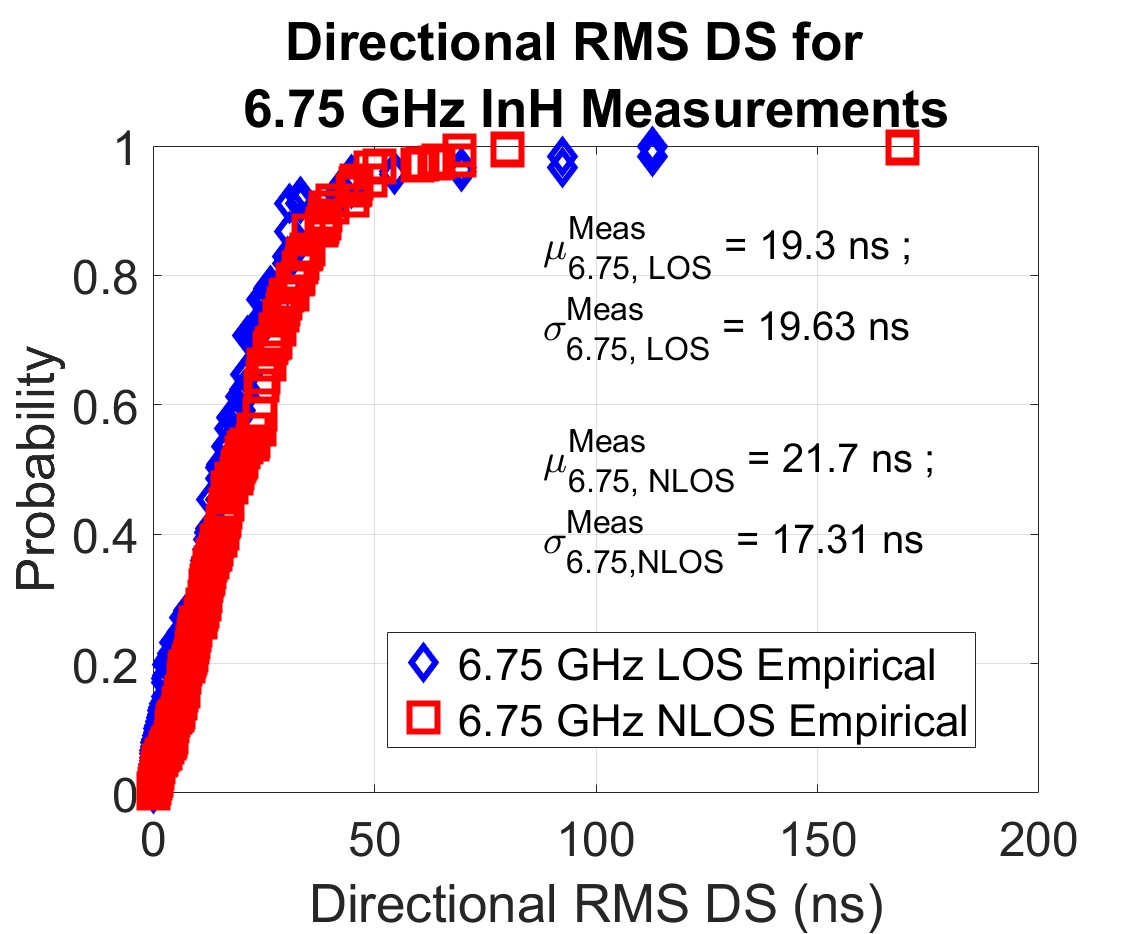}
			\label{fig:dirds7}
		}%
		\hfil 
		\subfloat[]{%
			\includegraphics[width=0.48\columnwidth]{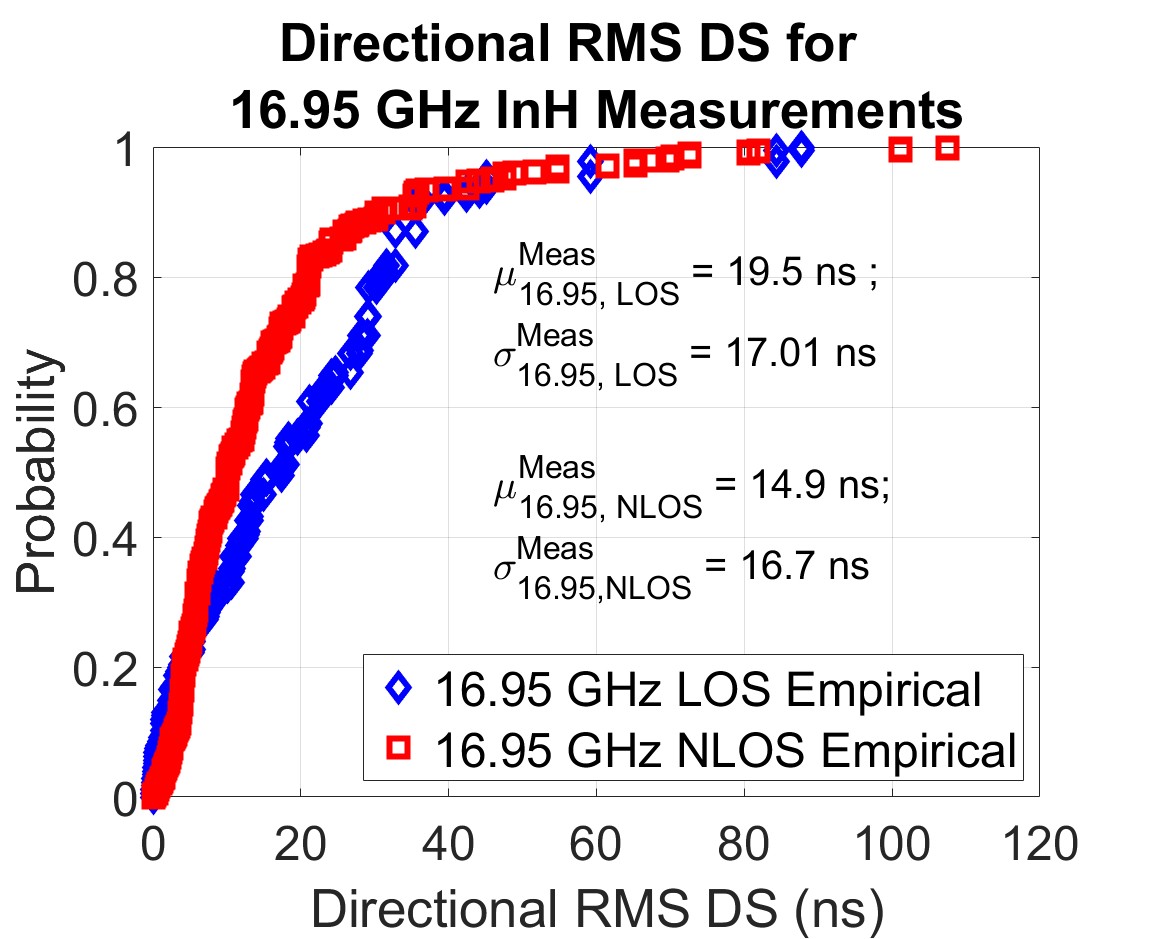}
			\label{fig:dirds17}
		}%
		\\ 
		\vspace{-10 pt}
		\subfloat[]{%
			\includegraphics[width=0.48\columnwidth]{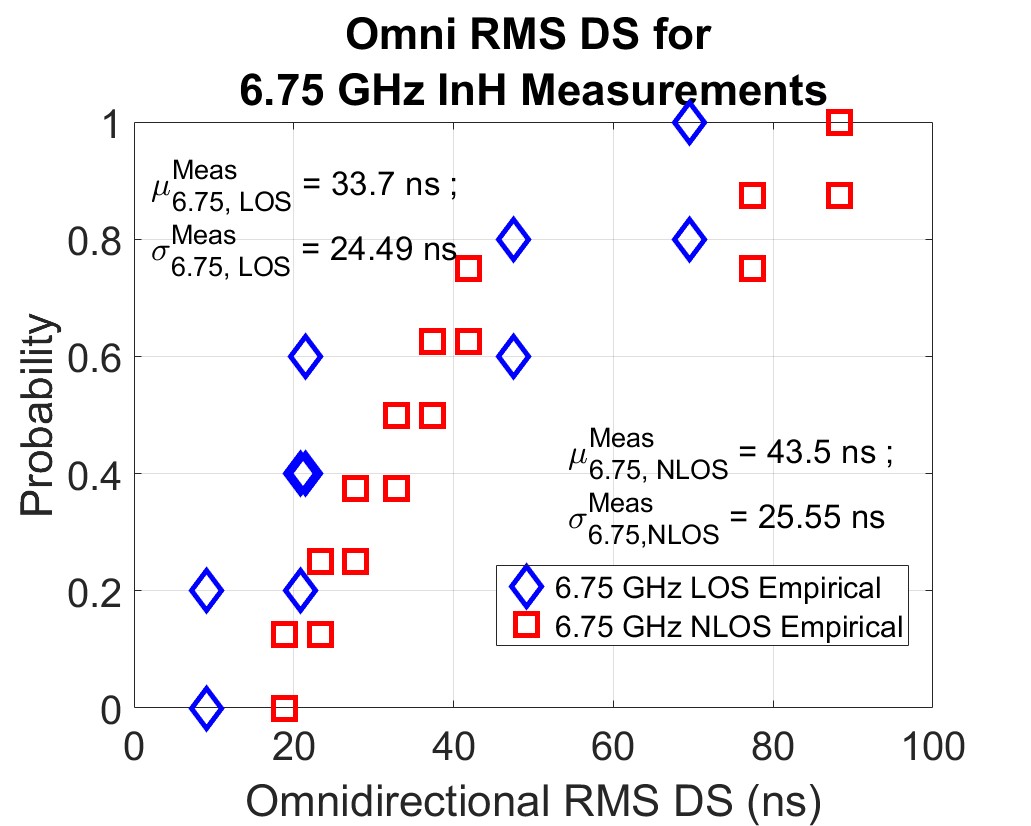}
			\label{fig:omnids7}
		}%
		\hfil 
		\subfloat[]{%
			\includegraphics[width=0.48\columnwidth]{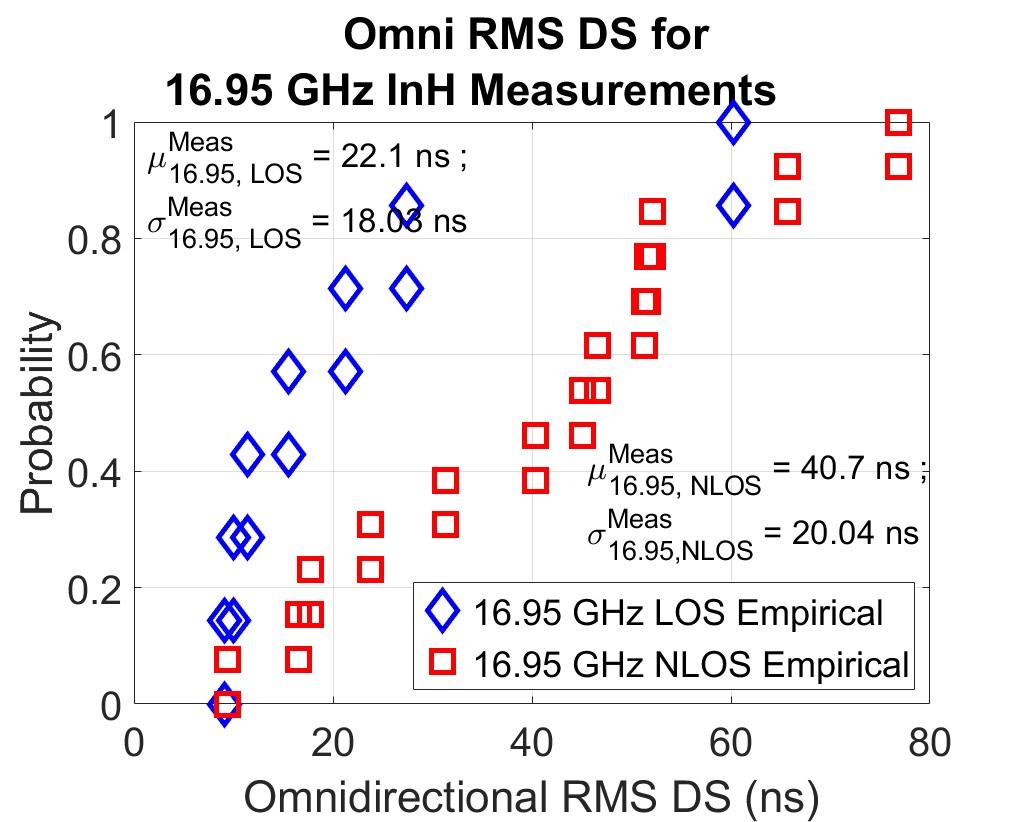}
			\label{fig:omnids17}
		}%
		\caption{InH LOS and NLOS RMS DSs at different frequencies: (a) Directional at 6.75 GHz; (b) Directional at 16.95 GHz; (c) Omnidirectional at 6.75 GHz; (d) Omnidirectional at 16.95 GHz. }
		\vspace{-5 pt}
		\label{fig:2x2fig}
	\end{figure}

	\renewcommand{\arraystretch}{1.2}
	\begin{table}[!t]
		\centering
		\caption{RMS Delay Spread Characteristics from 6.75 GHz to 142 GHz for LOS and NLOS Conditions.}
		\label{tab:RMS_DS}
		\begin{tabular}{|l|c|c|c|c|c|}
			\hline
			\textbf{Frequency (GHz)} & \textbf{6.75} & \textbf{16.95} & \textbf{28\cite{Ju2021jsac}} & \textbf{73\cite{Xing2021_Inicl}} & \textbf{142\cite{Xing2021_Inicl}} \\ \hline
			\multicolumn{6}{|c|}{Dir RMS DS}     \\ \hline
			LOS [\(\mu\) (ns)] & 19.3              & 19.5               & 3.9            & 3.5             & 2.7              \\ \hline
			NLOS [\(\mu\) (ns)]& 21.7              & 14.9               & 14.5            & 10.0            & 7.2              \\ \hline
			\multicolumn{6}{|c|}{Omni RMS DS}          \\ \hline
			LOS [\(\mu\) (ns)] & 33.7              & 22.1               & 10.8            & 6.2             & 3.0              \\ \hline
			NLOS [\(\mu\) (ns)]& 43.5              & 40.7               & 17.1            & 12.3            & 9.2              \\ \hline
		\end{tabular}
	\vspace{-1\baselineskip}
	\vspace{-10 pt}
	\end{table}
	\renewcommand{\arraystretch}{1}

	\section{Conclusion} \label{sxn:Conc}
	
	This paper reported comprehensive indoor propagation measurement campaigns for the FR1(C) and FR3 frequency bands in the indoor office/labs at the NYU WIRELESS Research Center. The results from the omnidirectional and directional CI PL modeling with a 1 m free space reference distance showed a waveguiding effect in the indoor hallways with omni PLEs of 1.40 and 1.32 in LOS at 6.75 and 16.95 GHz, respectively. A decrease in omnidirectional and directional RMS DS and azimuthal angular spread with increasing frequency was observed when compared with mmWave frequencies, indicating more multipath-rich propagation at the FR3 and FR1(C) frequencies. PAS showed widespread multipath arrival and departure directions, and spatial lobe and omnidirectional RMS AS was found narrower at 16.95 GHz compared to 6.75 GHz. The empirical data and models generated provide insights for future indoor propagation modeling in the 3GPP frequency bands for 5G and 6G and support the analysis and design of wireless systems and networks.

	\section*{Acknowledgment}
	Authors thank Profs. Sundeep Rangan and Marco Mezzavilla for support with licensing and Tomoki Yoshimura from Sharp Labs of America for support in the measurements.
	
	\bibliographystyle{IEEEtran}
	\bibliography{references}
	
\end{document}